\documentclass[11pt]{article}

\usepackage[a4paper,margin=1in]{geometry}
\usepackage[T1]{fontenc}
\usepackage{amsmath,amssymb}
\usepackage{booktabs}
\usepackage{microtype}
\usepackage{url}
\usepackage[hidelinks]{hyperref}
\usepackage{xcolor}
\usepackage{tikz}
\usetikzlibrary{arrows.meta,positioning,fit,backgrounds}

\newcommand{\smtwenty}{\texttt{sm\_20}}
\newcommand{\tps}{tok/s}

\title{\bf A 35B Hybrid-Attention Mixture-of-Experts Model on a 6\,GB 2011 GPU:\\
Hand-Written 4-bit CUDA Inference for Fermi (\smtwenty)}

\author{%
  A. C. Opus, J. Q. Lu\thanks{Correspondence: \texttt{junqiang.lu@upr.edu}.}\\
  \small Department of Physics, University of Puerto Rico, Mayaguez, PR 00680, USA
}
\date{June 2026}

\begin{document}
\maketitle

\begin{abstract}
We report end-to-end inference of \textbf{Qwen3.6-35B-A3B}---a 35-billion-parameter,
$\sim$3B-active Mixture-of-Experts (MoE) model with a hybrid gated-delta-net /
full-attention backbone---on a \textbf{2011 NVIDIA Tesla C2075} (Fermi, compute
capability \smtwenty, 6\,GB), a GPU that predates tensor cores, native FP16
arithmetic, the \texttt{DP4A} integer dot-product instruction, and support in
every modern CUDA toolchain. Because the 4-bit model ($\approx$10.5\,GB) is
roughly twice the device memory, we adopt a \emph{hybrid} execution strategy:
the GPU performs batched prompt \emph{prefill} with expert weights streamed
layer-by-layer from host RAM, while \emph{decode} runs on the host CPU using a
hand-written W4A8 integer GEMV built on the SSSE3 \texttt{pmaddubsw} instruction.
The entire engine---GEMM, hybrid-attention recurrence, MoE routing, and a
from-scratch vision tower---is written by hand for \smtwenty{} and compiled with
the legacy CUDA 8.0 toolchain. On a 947-token prompt we reduce prefill latency
from 57.2\,s to 37.5\,s ($-34\%$) through expert pinning, single-pass prefill,
and NUMA interleaving, and we raise decode throughput from 2.8 to 8.6\,\tps{}
($\approx 3\times$) with the integer-SIMD kernel. A position-indexed snapshot
cache for the recurrent (gated-delta-net) state restores prefix reuse on a
recurrent architecture, cutting a repeated 78\,s prefill to 0.5\,s. We also
report a set of \emph{negative} results---offloading the language-model head to
the idle GPU, hyper-threading, and three GPU-kernel rewrites all fail to help---%
which together pin down the practical floor of this hardware. Our aim is not a
speed record but a careful account of what it takes, and where the walls are,
to run a contemporary frontier-class MoE on fourteen-year-old silicon.
\end{abstract}

\section{Introduction}

The hardware demands of large language models are usually framed as a frontier
problem: newer accelerators, more memory, faster interconnects. This paper asks
the opposite question. \emph{How far down the hardware stack can a modern
Mixture-of-Experts (MoE) model be pushed, and what exactly breaks along the
way?} Concretely, we take a 35-billion-parameter MoE with a hybrid
linear/full-attention backbone and run it, in 4-bit, on an NVIDIA Tesla C2075:
a Fermi-generation GPU released in 2011 with 6\,GB of GDDR5, 448 CUDA cores, a
peak of roughly 1\,TFLOP of FP32, and---critically---\emph{none} of the
primitives that modern inference stacks assume. It has no tensor cores
(introduced with Volta), no high-throughput FP16 arithmetic, and no \texttt{DP4A}
4-element 8-bit integer dot product (introduced with Pascal, \texttt{sm\_61}).
Its compute capability, \smtwenty, was dropped from the CUDA toolkit years ago;
the last toolchain that targets it is CUDA 8.0.

This combination of constraints makes almost every off-the-shelf component
unusable. Modern kernels assume \texttt{mma} tensor-core intrinsics or at least
\texttt{DP4A}; modern toolchains will not emit \smtwenty{} code; and the model
itself---roughly 10.5\,GB even at 4 bits per weight---does not fit in 6\,GB of
device memory. We therefore build the inference engine from first principles. It
is a fork of a minimal single-file CUDA port~\cite{llama2cu} of Karpathy's
\texttt{llama2.c}~\cite{llama2c}, extended into a complete MoE engine with a
hand-written 4-bit GEMM, a hybrid-attention recurrence kernel, host-side
expert offload, and a multimodal vision path. The contributions are:

\begin{enumerate}
\item \textbf{A working hybrid GPU-prefill / CPU-decode engine for \smtwenty}.
  Expert weights live in host RAM; the GPU streams them layer-by-layer for
  batched prefill, and the CPU computes single-token decode with a hand-written
  W4A8 integer kernel. The split is dictated by the device-memory ceiling and
  the absence of fast integer GPU dot products.
\item \textbf{An integer-SIMD decode path}. We pack 4-bit weights and 8-bit
  activations into the SSSE3 \texttt{pmaddubsw} multiply-add, mirroring the
  $\mathrm{Q4\_0}\times\mathrm{Q8\_0}$ scheme of \texttt{llama.cpp}~\cite{llamacpp},
  on a host CPU that lacks AVX. This takes decode from 2.8 to 8.6\,\tps{}.
\item \textbf{A prefix cache for a recurrent architecture}. Gated-delta-net
  state is only valid at the end of a sequence, which defeats naive
  prefix-KV reuse. We take position-indexed boundary snapshots of the recurrent
  state during both prefill and decode, enabling partial reuse; a repeated
  prompt's prefill drops from 78\,s to 0.5\,s.
\item \textbf{Prefill optimizations and an honest hardware floor}. Expert
  pinning, single-pass prefill, and NUMA interleaving cut prefill by
  $\sim$35\%. We then show, with measurements, that the remaining cost is
  structural: a thin-$M$ MoE GEMM and an inherently sequential recurrence sit
  near the practical limit of Fermi, and three further kernel rewrites,
  GPU-offloading the LM head, and hyper-threading all fail to help.
\item \textbf{Multimodality}. A from-scratch vision tower for \smtwenty{}
  reproduces a reference implementation to $\sim$$4\times10^{-4}$ per element
  and correctly discriminates simple shapes and colors.
\end{enumerate}

We emphasize that this is an engineering and measurement study on a single
machine, not a claim of a new algorithm or a throughput record. The value is in
the constraints: each one forces a design decision, and the negative results
are as informative as the positive ones.

\section{Background}

\subsection{The target hardware}
The NVIDIA Tesla C2075 is a Fermi-architecture (GF110) compute card from 2011:
448 CUDA cores, 6\,GB GDDR5 ($\approx$5.2\,GB usable with ECC) at $\sim$144\,GB/s,
$\sim$1.03\,TFLOP FP32 peak, compute capability \smtwenty, and a PCIe~2.0 host
link. Relative to any accelerator a modern stack targets, it lacks tensor
cores, fast FP16, \texttt{DP4A}, and asynchronous-copy/residency features; it is
limited to 48\,KB of shared memory per block and a 65{,}535 grid-dimension cap.
The host is a dual-socket Intel Xeon X5690 (Westmere-EP, $2\times6$ physical
cores / 24 threads, 3.47\,GHz) with two NUMA nodes and DDR3 memory; notably it
predates AVX, so host SIMD is limited to SSE4.2/SSSE3. We compile device code
with CUDA 8.0---the last toolkit supporting \smtwenty---inside a container, and
the host decode loops with a modern GCC restricted to \texttt{-march=westmere}.

\subsection{The model}
Qwen3.6-35B-A3B belongs to the Qwen3 series~\cite{qwen3}. It is a sparse MoE
with $\sim$35B total and $\sim$3B active parameters: 40 transformer blocks, each
with a sparsely-gated MoE feed-forward layer~\cite{shazeer2017} (256 experts,
top-8 routing, plus a shared always-on expert) and a small per-block hidden
width. Its backbone is \emph{hybrid}: 30 of the 40 blocks use a gated-delta-net
(GDN) linear-attention layer~\cite{gateddeltanet}---a gated extension of the
delta rule in the state-space/linear-attention lineage of
Mamba~\cite{mamba}---and the remaining 10 use standard softmax
attention~\cite{flashattention}. The GDN layer carries a recurrent state
$S\in\mathbb{R}^{H_k\times H_v}$ per value head that is updated token-by-token; in
our configuration there are 32 value heads with $H_k=H_v=128$. This recurrence
is the source of both the architecture's efficiency and, as we show, its
prefix-caching difficulty.

\subsection{Weight quantization}
Weight-only 4-bit quantization is now standard for LLM
deployment~\cite{gptq,awq}. We use a group-wise symmetric 4-bit format in the
$\mathrm{Q4\_0}$ family~\cite{llamacpp}: each weight is an integer in
$\{-7,\dots,7\}$ with one FP32 scale per group, packed two values per byte. For decode we pair
4-bit weights with 8-bit activations (W4A8) so the inner product can be issued
as an integer multiply-add. The active model after pruning (below) is
$\approx$10.5\,GB on disk, about twice the device memory.

\subsection{Expert pruning}
To shrink the resident footprint we apply REAP (Router-weighted Expert
Activation Pruning)~\cite{reap}, a one-shot MoE compression method that ranks
experts by a combination of router gate-values and expert activation norms and
\emph{removes} (rather than merges) the least useful ones. We use a 50\%
pruning ratio, taking the model from 256 to 128 experts per layer. REAP reports
near-lossless behavior at 50\% on generative tasks~\cite{reap}; consistent with
this, our downstream task accuracy is unchanged after pruning (Section~\ref{sec:eval}).

\section{System Design}

\subsection{Why hybrid GPU-prefill / CPU-decode}
\begin{figure}[t]
\centering
\begin{tikzpicture}[font=\small,>=Stealth,
  box/.style={draw,rounded corners,align=center,minimum height=8mm,fill=white,inner sep=3pt},
  reg/.style={draw,rounded corners,inner sep=3.5mm}]
  \node[box,minimum width=4.3cm] (exp) {4-bit experts $\sim$9\,GB\\(page-locked, NUMA-interleaved)};
  \node[box,minimum width=4.3cm,below=8mm of exp] (cpu) {CPU thread pool\\W4A8 \texttt{pmaddubsw} GEMV};
  \begin{scope}[on background layer]
    \node[reg,fill=blue!4,fit=(exp)(cpu),
      label={[font=\small]above:\textbf{Host} (2$\times$Xeon, DDR3, 2 NUMA nodes)}] (host) {};
  \end{scope}
  \node[box,minimum width=4.3cm,right=3.6cm of exp] (scr) {per-layer expert scratch};
  \node[box,minimum width=4.3cm,below=4mm of scr] (gem) {grouped 4-bit GEMM};
  \node[box,minimum width=4.3cm,below=4mm of gem] (ker) {GDN recurrence $+$ attention};
  \begin{scope}[on background layer]
    \node[reg,fill=green!5,fit=(scr)(gem)(ker),
      label={[font=\small]above:\textbf{GPU} (Tesla C2075, 6\,GB, \smtwenty)}] (gpu) {};
  \end{scope}
  \draw[->,thick] (exp.east) -- node[above,font=\scriptsize,align=center]{\textbf{prefill}\\stream / layer}
                                 node[below,font=\scriptsize]{PCIe 2.0} (scr.west);
  \draw[->,thick] (scr) -- (gem);
  \draw[->,thick] (gem) -- (ker);
  \draw[->,thick,dashed] (cpu.west) .. controls +(-1.4,0) and +(-1.4,0) ..
        node[left,font=\scriptsize,align=center]{\textbf{decode}\\(host-resident,\\1 token / step)} (exp.west);
\end{tikzpicture}
\caption{Hybrid execution. The 4-bit experts reside in page-locked,
NUMA-interleaved host memory. \emph{Prefill} streams each layer's experts across
PCIe~2.0 into a 6\,GB-resident scratch buffer and runs the batched GPU kernels;
\emph{decode} keeps the single-token computation on the host CPU (W4A8 integer
SIMD), avoiding a per-token PCIe round trip. Pruning and the prefix cache act on
the prefill (GPU) path; the integer kernel acts on the decode (CPU) path.}
\label{fig:system}
\end{figure}
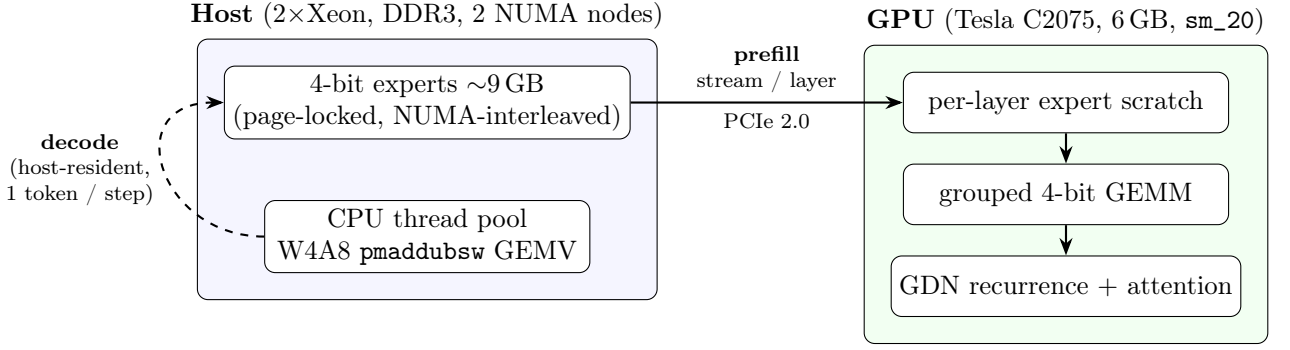

Two facts determine the architecture. First, the experts do not fit in device
memory, so they reside in host RAM and are moved to the GPU on demand. Second,
the cost of that movement is amortized very differently between the two phases
of inference. During \emph{prefill} the model processes hundreds of prompt
tokens at once, so streaming a layer's experts into a 6\,GB-resident scratch
buffer and running a batched GEMM pays off: the host$\to$device transfer is
shared across all tokens in the chunk. During \emph{decode} only one token is
produced per step; streaming a layer's experts across PCIe~2.0 to multiply a
single activation vector is dominated by transfer latency. We therefore keep
decode on the host CPU, reading the 4-bit experts directly from RAM. The
production binary is, accordingly, a \emph{GPU-prefill, CPU-decode} engine
(Figure~\ref{fig:system}).

\subsection{GPU prefill}
For each layer the engine (i) streams that layer's expert block host$\to$device,
(ii) runs the GDN recurrence or softmax-attention kernel, and (iii) runs the MoE
feed-forward as a single \emph{grouped} 4-bit GEMM over all routed
token--expert pairs. The GEMM is a standard register-tiled kernel
($64\times64$ output tiles, $4\times4$ register micro-tiles, 256 threads/block)
that dequantizes the 4-bit weight tile into shared memory before accumulation;
grouping all experts into one launch avoids the low occupancy of one launch per
expert. Attention uses an online-softmax formulation in the spirit of
FlashAttention~\cite{flashattention} but written for \smtwenty{} (no tensor
cores, no \texttt{cp.async}). The GDN recurrence is a custom kernel that carries
the per-head state through the token loop.

\subsection{CPU decode: W4A8 with \texttt{pmaddubsw}}
On the host, decode is a pthread pool (one thread per physical core) running a
W4A8 GEMV. Activations are quantized per group to 8-bit with an FP32 scale; each
4-bit weight is mapped to an unsigned value in $\{1,\dots,15\}$, and the dot product is
accumulated with the SSSE3 \texttt{pmaddubsw} instruction, which performs sixteen
unsigned$\times$signed 8-bit multiply-adds per issue. This is the same
arithmetic \texttt{llama.cpp}~\cite{llamacpp} uses for
$\mathrm{Q4\_0}\times\mathrm{Q8\_0}$, specialized here for a CPU without AVX.
Replacing the scalar FP32 inner loop with this integer-SIMD kernel is the single
largest decode win (Section~\ref{sec:eval}).

\subsection{Resident process and prefix cache}
\label{sec:cache}
Earlier iterations reloaded the 10.5\,GB model per request; we instead keep the
model resident and serve requests over a pipe, which removes a $\sim$15\,s
per-request reload. On top of this we add a prefix cache. The difficulty is the
GDN recurrence: unlike a softmax-attention KV cache, which is valid at every
position, the recurrent state is only meaningful at the end of the processed
sequence, so a new prompt cannot simply reuse a prefix unless it continues the
old one exactly. We borrow the idea of \emph{boundary snapshots}: we checkpoint
the full recurrent state at block boundaries during both prefill and decode, and
index these snapshots by absolute token position. A new request scans for the
largest snapshot whose position is a prefix of the request and resumes from
there. Because the snapshots are position-indexed (not tied to a fixed grid),
the same mechanism supports both exact continuation (multi-turn) and partial
prefix reuse, degrading gracefully to a full recompute on a miss.

\section{Optimizations}

\subsection{Expert pinning}
The expert block is moved from a pageable file mapping into page-locked host
memory (a single $\sim$9\,GB pinned allocation), which roughly halves the
effective host$\to$device bandwidth cost: the per-prompt expert-transfer time
falls from 16.0\,s to 6.4\,s. (Page-locking the file mapping directly is
rejected by the driver, so the experts are copied once into a pinned buffer at
load time.)

\subsection{Single-pass prefill}
\label{sec:singlechunk}
The boundary-snapshot machinery originally chunked prefill into 256-token
segments, which had an unintended cost: each segment re-streamed the full expert
set and ran the GEMM at a small batch size. Processing the whole prompt in one
pass eliminates the redundant streaming and raises GEMM occupancy. The effect is
large and is the dominant prefill win; Table~\ref{tab:chunk} shows prefill
latency as a function of segment size on a 947-token prompt. We retain
position-indexed snapshots (Section~\ref{sec:cache}) so that single-pass prefill
does not sacrifice prefix reuse.

\begin{table}[t]
\centering
\caption{Prefill latency vs.\ prefill segment size (947-token prompt). One pass
($\geq$prompt length) eliminates per-segment expert re-streaming and maximizes
GEMM batch size.}
\label{tab:chunk}
\begin{tabular}{lccc}
\toprule
Segment size (tokens) & 256 & 512 & $\geq$947 (single pass)\\
\midrule
Prefill latency (s)   & 51.9 & 40.7 & \textbf{35.0}\\
MoE GEMM share (s)    & 32.6 & 22.1 & \textbf{16.9}\\
Expert transfer (s)   & 7.2  & 3.6  & \textbf{1.8}\\
\bottomrule
\end{tabular}
\end{table}

\subsection{NUMA interleaving}
The host is a two-socket NUMA machine. Allocating the pinned experts on a single
node forces half the decode threads to read across the inter-socket link. We set
an interleaved memory policy at startup so the expert block and the weight
mapping are spread across both memory controllers, which yields a small but free
decode improvement ($\sim$5\%). The modest size of this effect is itself
informative: decode is not bottlenecked on cross-socket bandwidth.

\subsection{Pruning as a prefill, not decode, lever}
REAP halves the expert count, which shrinks load and prefill time (fewer experts
to stream and to route over). It does \emph{not} change decode latency: decode
reads only the top-8 routed experts per layer, a quantity independent of the
total expert count. We confirm this empirically---decode throughput is identical
at 256 and 128 experts---and it clarifies that pruning and the decode kernel are
orthogonal levers.

\section{Evaluation}
\label{sec:eval}

All measurements are on the single C2075 workstation described above, using the
REAP-pruned 128-expert model in the group-wise 4-bit format, with the W4A8
decode kernel enabled.

\subsection{Latency}
\begin{figure}[t]
\centering
\begin{tikzpicture}[font=\small]
  \def\s{0.34} 
  \foreach \x in {0,5,10,15} {
    \draw[gray!25] (\x*\s,-0.15) -- (\x*\s,2.25);
    \node[below,font=\scriptsize,text=gray] at (\x*\s,-0.15) {\x};
  }
  \fill[gray!45] (0,0) rectangle (7.0*\s,0.55);
  \node[left,font=\scriptsize] at (-0.1,0.275) {attention};
  \node[right,font=\scriptsize] at (7.0*\s,0.275) {7.0\,s};
  \fill[gray!65] (0,0.75) rectangle (11.9*\s,1.30);
  \node[left,font=\scriptsize] at (-0.1,1.025) {GDN recurrence};
  \node[right,font=\scriptsize] at (11.9*\s,1.025) {11.9\,s};
  \fill[blue!25] (0,1.50) rectangle (1.8*\s,2.05);
  \fill[blue!55] (1.8*\s,1.50) rectangle (17.8*\s,2.05);
  \node[left,font=\scriptsize] at (-0.1,1.775) {MoE};
  \node[right,font=\scriptsize] at (17.8*\s,1.775) {17.8\,s};
  \fill[blue!25] (0,-0.95) rectangle +(0.32,0.32);
  \node[right,font=\scriptsize] at (0.4,-0.79) {expert transfer (1.8\,s)};
  \fill[blue!55] (4.2,-0.95) rectangle +(0.32,0.32);
  \node[right,font=\scriptsize] at (4.6,-0.79) {MoE 4-bit GEMM (16.0\,s)};
  \node[below,font=\scriptsize,text=gray] at (8*\s,-1.35) {prefill time (s)};
\end{tikzpicture}
\caption{Where prefill time goes (single-pass, 947-token prompt; all three are
GPU kernels). The MoE GEMM dominates and, after single-pass prefill, carries only
1.8\,s of expert host$\to$device transfer. The GDN recurrence (a sequential scan)
and softmax attention account for the rest; the three components sum to
$\approx$36.7\,s of the 37.5\,s prefill.}
\label{fig:breakdown}
\end{figure}
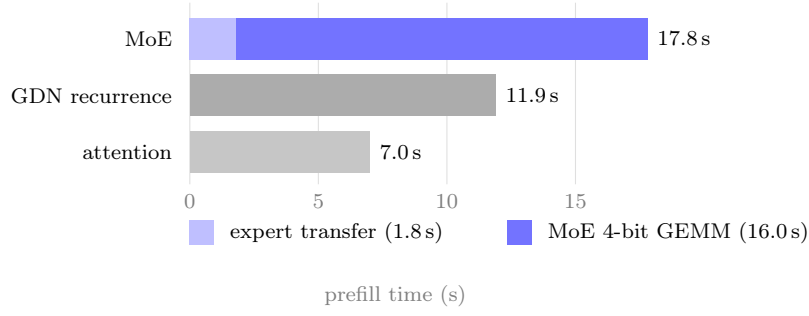

Table~\ref{tab:main} summarizes prefill and end-to-end latency before and after
the optimization stack (pinning $+$ single-pass prefill $+$ NUMA interleaving).
Prefill falls by 34--36\%. On a 947-token prompt, a single forward pass spends
17.8\,s in the MoE GEMM (47\%), 11.9\,s in the GDN recurrence (31\%), and 7.0\,s
in softmax attention (18\%) (Figure~\ref{fig:breakdown}); all three are GPU kernels, and after single-pass
prefill the expert transfer is only 1.8\,s. Prefill therefore dominates
end-to-end latency for these prompt lengths, and the system is, in steady state,
a GPU-bound prefill engine feeding a CPU-bound decoder.

\begin{table}[t]
\centering
\caption{Prefill and end-to-end latency, before vs.\ after the
pinning $+$ single-pass $+$ NUMA optimization stack. ``e2e'' includes 60
generated tokens. Prompt lengths are 947 (``1K'') and 1{,}871 (``2K'') tokens.}
\label{tab:main}
\begin{tabular}{lcccc}
\toprule
 & \multicolumn{2}{c}{Prefill (s)} & \multicolumn{2}{c}{End-to-end (s)}\\
\cmidrule(lr){2-3}\cmidrule(lr){4-5}
Prompt & before & after & before & after\\
\midrule
1K (947 tok)   & 57.2 & \textbf{37.5} ($-34\%$) & 63.7 & \textbf{45.1}\\
2K (1{,}871 tok) & 122.9 & \textbf{78.4} ($-36\%$) & 130.1 & \textbf{86.5}\\
\bottomrule
\end{tabular}
\end{table}

\subsection{Decode throughput}
The W4A8 integer-SIMD kernel takes decode from a scalar-FP32 baseline of
2.8\,\tps{} to $\sim$6.1\,\tps{}, and NUMA interleaving brings it to
$\sim$8.6\,\tps{}---roughly a $3\times$ improvement end-to-end on the decode
path. A per-component decode profile shows that the routed experts account for
only about a quarter of decode time (reducing top-$k$ from 8 to 4 changes
per-token latency from 116 to 103\,ms); the remainder is the dense per-layer
projections and the language-model head, read from host RAM. This is why pruning
does not help decode and why lower-bit \emph{experts} alone would not move it
much.

\subsection{Prefix reuse}
With the position-indexed snapshot cache, a request that shares a long prefix
with a prior one resumes from the nearest snapshot instead of recomputing. On a
repeated 1{,}871-token prompt, prefill drops from 78.4\,s to 0.5\,s and
end-to-end latency from 86.5\,s to 6.1\,s. Reuse requires the shared prefix to
reach a snapshot boundary; short prompts and certain multi-turn chat templates
(which rewrite earlier turns) miss and fall back to a full pass, losslessly.

\subsection{Correctness and multimodality}
Outputs are coherent and task-correct: on a small held-out exam-grading task the
pruned 4-bit model reproduces the reference rubric score, and the W4A8 path is
bit-stable against the scalar path on short generations (it diverges only on
long greedy chains, as expected from accumulated low-bit activation error). The
engine is also multimodal: we port a vision-transformer tower to \smtwenty{} for
a 9B vision-language sibling model, reproducing the reference image embeddings to
$\sim$$4\times10^{-4}$ per element and correctly distinguishing, e.g., a red
circle from a blue square via image-embedding injection into the language model.

\subsection{Negative results: the hardware floor}
A central finding is what \emph{does not} work, because it locates the practical
limit of the platform.
\begin{itemize}
\item \textbf{Offloading the LM head to the (idle) GPU is slower}, not faster:
  $116\to139$\,ms/token. Although the GPU is idle during decode, a per-token
  round trip over PCIe~2.0 plus a Fermi-speed GEMM costs more than the host
  integer kernel it replaces.
\item \textbf{Hyper-threading is catastrophic.} Using all 24 logical cores
  instead of 12 physical cores raises per-token latency from 116 to
  1{,}722\,ms, as the busy-wait thread pool oversubscribes each physical core's
  issue ports. Twelve threads (one per physical core) is optimal.
\item \textbf{Three prefill-kernel rewrites do not help.} Splitting the GDN
  recurrence across more blocks for occupancy is neutral; moving the recurrent
  state into shared memory is slightly slower (the larger shared-memory
  footprint reduces occupancy more than it saves on global traffic, because the
  Fermi caches already serve the per-head state well); and enlarging the GEMM
  $K$-tile is slightly slower. The MoE GEMM runs well below FP32 peak because it
  is a \emph{thin}-$M$ problem ($\approx$59 tokens per expert), and the GDN
  layer is an inherently sequential recurrence. Both are structural, not tuning,
  limits.
\end{itemize}
Pushing prefill substantially below the reported $-35\%$ would require
algorithmic changes---a chunked/parallel scan for the recurrence, or a different
MoE batching strategy---rather than kernel tuning. We also did not pursue
multi-token (speculative) decoding: at batch size one on a sparse MoE, the
expert-loading overhead makes it net-negative, consistent with reports for this
class of model.

\section{Related Work}
Our engine descends from \texttt{llama2.c}~\cite{llama2c} and its CUDA
port~\cite{llama2cu}, and adopts the 4-bit weight / 8-bit activation arithmetic
of \texttt{llama.cpp}~\cite{llamacpp}, whose CPU-with-GPU-offload design is the
closest in spirit to ours. The model follows the Qwen3 series~\cite{qwen3} and
its sparsely-gated MoE layers~\cite{shazeer2017}; its linear-attention blocks are
gated-delta-net layers~\cite{gateddeltanet} in the state-space/linear-attention
line opened by Mamba~\cite{mamba}, and its softmax-attention blocks are
computed with an online-softmax kernel in the FlashAttention
style~\cite{flashattention}. Weight-only 4-bit quantization follows the
now-standard PTQ literature~\cite{gptq,awq}, and expert pruning uses
REAP~\cite{reap}. Our contribution is not any of these techniques individually
but their composition under an extreme hardware constraint, together with the
recurrent-state prefix cache and the measured account of the platform's floor.

\section{Conclusion}
A 35B hybrid-attention MoE---a 2026-class frontier architecture---can be made to
run, in 4-bit, with both text and vision, on a 6\,GB GPU from 2011, by writing
the entire engine by hand for an unsupported compute capability and splitting
work between a streaming GPU prefill and an integer-SIMD CPU decode. Careful
systems work (expert pinning, single-pass prefill, NUMA interleaving, a
recurrent-state prefix cache) cuts prefill by about a third and triples decode,
and an equally careful set of negative results shows where the silicon's
ceiling lies. Beyond the novelty, the exercise is a reminder that the
assumptions baked into modern inference stacks---tensor cores, fast integer dot
products, device memory that holds the model---are conveniences, not
requirements, and that a surprising amount of contemporary capability is
reachable on hardware that the software ecosystem abandoned long ago.

\section*{Artifact and reproducibility}
The engine is a single CUDA/C source compiled with CUDA 8.0 for \smtwenty{} (in
a container) plus a host decode object compiled with GCC under
\texttt{-march=westmere}. All experiments run on one workstation; we report
wall-clock latency and tokens/second directly, and avoid derived FLOP-efficiency
figures that depend on undisclosed internal dimensions. No hostnames, addresses,
or file paths are part of the artifact.


\begin{thebibliography}{99}

\bibitem{qwen3}
A.~Yang, A.~Li, et al. (Qwen Team, Alibaba).
\newblock Qwen3 Technical Report.
\newblock arXiv:2505.09388, 2025.

\bibitem{gateddeltanet}
S.~Yang, J.~Kautz, and A.~Hatamizadeh.
\newblock Gated Delta Networks: Improving Mamba2 with Delta Rule.
\newblock arXiv:2412.06464, 2024. (ICLR 2025.)

\bibitem{mamba}
A.~Gu and T.~Dao.
\newblock Mamba: Linear-Time Sequence Modeling with Selective State Spaces.
\newblock arXiv:2312.00752, 2023.

\bibitem{shazeer2017}
N.~Shazeer, A.~Mirhoseini, K.~Maziarz, A.~Davis, Q.~Le, G.~Hinton, and J.~Dean.
\newblock Outrageously Large Neural Networks: The Sparsely-Gated
  Mixture-of-Experts Layer.
\newblock arXiv:1701.06538, 2017.

\bibitem{reap}
M.~Lasby, I.~Lazarevich, N.~Sinnadurai, S.~Lie, Y.~Ioannou, and V.~Thangarasa.
\newblock REAP the Experts: Why Pruning Prevails for One-Shot MoE Compression.
\newblock arXiv:2510.13999, 2025. (ICLR 2026.)

\bibitem{gptq}
E.~Frantar, S.~Ashkboos, T.~Hoefler, and D.~Alistarh.
\newblock GPTQ: Accurate Post-Training Quantization for Generative Pre-trained
  Transformers.
\newblock arXiv:2210.17323, 2022. (ICLR 2023.)

\bibitem{awq}
J.~Lin, J.~Tang, H.~Tang, S.~Yang, X.~Dang, C.~Gan, and S.~Han.
\newblock AWQ: Activation-aware Weight Quantization for LLM Compression and
  Acceleration.
\newblock arXiv:2306.00978, 2023. (MLSys 2024.)

\bibitem{flashattention}
T.~Dao, D.~Y.~Fu, S.~Ermon, A.~Rudra, and C.~R\'e.
\newblock FlashAttention: Fast and Memory-Efficient Exact Attention with
  IO-Awareness.
\newblock arXiv:2205.14135, 2022.

\bibitem{llama2c}
A.~Karpathy.
\newblock llama2.c: Inference Llama 2 in one file of pure C.
\newblock \url{https://github.com/karpathy/llama2.c}.

\bibitem{llamacpp}
G.~Gerganov and the \texttt{llama.cpp} contributors.
\newblock llama.cpp: LLM inference in C/C++.
\newblock \url{https://github.com/ggml-org/llama.cpp}.

\bibitem{llama2cu}
R.~Allen.
\newblock llama2.cu: A CUDA port of llama2.c.
\newblock \url{https://github.com/rogerallen/llama2.cu}.

\end{thebibliography}
\end{document}